\pdfoutput=1

\def\40K{$^{40}$K}

\def\K{$^{39}$K}
\def\Na{$^{23}$Na}
\def\NaK{\Na\K}

\def\ket#1{\mathinner{|{#1}\rangle}}

\documentclass[aps,prl,twocolumn,floatfix,10pt,superscriptaddress]{revtex4-1} 
\usepackage{graphicx,amsmath,amssymb,ulem}
\usepackage{textcomp}
\usepackage{hyperref}
\usepackage{color}
\usepackage{epstopdf}
\usepackage{nicefrac}
\usepackage[caption=false]{subfig}
\usepackage{gensymb}
\usepackage{placeins}
\begin{document}

\title{An Ultracold Gas of Bosonic \NaK\ Ground-State Molecules}

\author{Kai K.~Voges}
\email{voges@iqo.uni-hannover.de}
\affiliation{Institut f\"ur Quantenoptik, Leibniz Universit\"at Hannover, 30167~Hannover, Germany}
\author{Philipp~Gersema}
\affiliation{Institut f\"ur Quantenoptik, Leibniz Universit\"at Hannover, 30167~Hannover, Germany}
\author{Mara Meyer zum Alten Borgloh}
\affiliation{Institut f\"ur Quantenoptik, Leibniz Universit\"at Hannover, 30167~Hannover, Germany}
\author{\\Torben A.~Schulze}
\author{Torsten~Hartmann}
\affiliation{Institut f\"ur Quantenoptik, Leibniz Universit\"at Hannover, 30167~Hannover, Germany}
\author{Alessandro~Zenesini}
\affiliation{Institut f\"ur Quantenoptik, Leibniz Universit\"at Hannover, 30167~Hannover, Germany}
\affiliation{INO-CNR BEC Center and Dipartimento di Fisica, Universit\`{a} di Trento, 38123 Povo, Italy}
\author{Silke~Ospelkaus}
\email{silke.ospelkaus@iqo.uni-hannover.de}
\affiliation{Institut f\"ur Quantenoptik, Leibniz Universit\"at Hannover, 30167~Hannover, Germany}

\date{\today}

\begin{abstract}
We report the creation of ultracold bosonic dipolar \NaK\ molecules in their absolute rovibrational ground state. Starting from weakly bound molecules immersed in an ultracold atomic mixture, we coherently transfer the dimers to the rovibrational ground state using an adiabatic Raman passage. 
We analyze the  two-body decay in a pure molecular sample and in molecule-atom mixtures and find an unexpectedly low two-body decay coefficient for collisions between molecules and \K\ atoms in a selected hyperfine state. The preparation of bosonic \NaK\ molecules opens the way for future comparisons between fermionic and bosonic ultracold ground-state molecules of the same chemical species.
\end{abstract}
\maketitle
\begin{figure}[t]
	\includegraphics[width=1\columnwidth]{%Pictures/STIRAPDetuning/LDetuning.pdf}
		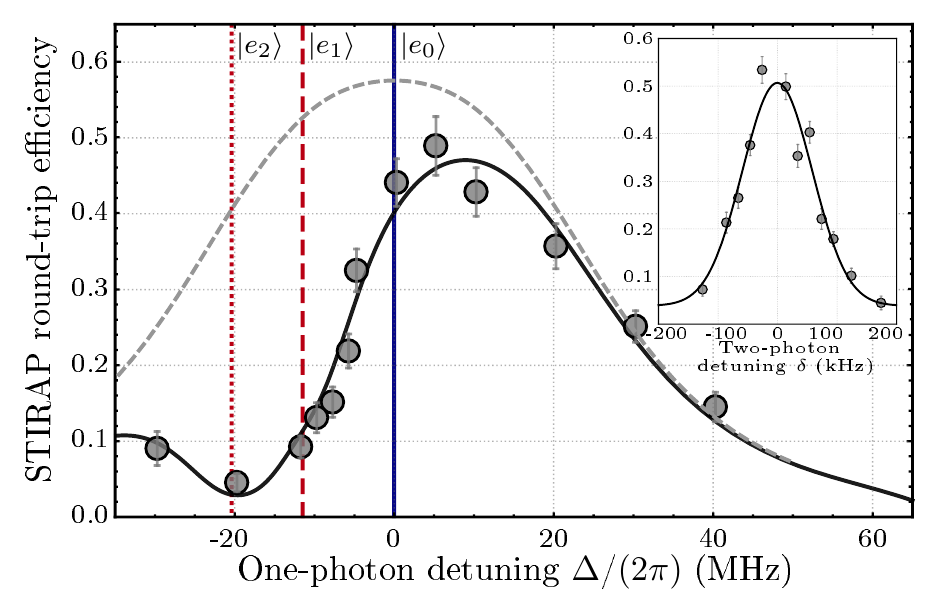}
	\caption{One- and two-photon detuning for STIRAP. The round-trip efficiency for STIRAP is shown as a function of the one-photon detuning $\Delta$. The pulse sequence and laser intensities for these measurements were kept constant corresponding to the optimal values given in the text. The vertical solid blue(dashed red)[dotted red] line is the position of the $\ket{e_{0(1)[2]}}$ state deduced from measurements and the model developed in \cite{voges2019pathway}. The solid black curve is a fit using the 5-level master equation model and the individual couplings of the Stokes laser to the $\ket{e_{1,2}}$ states as free parameters. The dashed gray curve is a theory curve from a 3-level model using the same set of parameters. The inset shows the STIRAP round-trip efficiency dependent on the two-photon detuning $\delta$ with a phenomenological Gaussian fit. The error bars for both plots are the standard error coming from different experimental cycles.}
	\label{SingPhotDetPic}
\end{figure}
\begin{figure}[t]
	\includegraphics[width=1\columnwidth]{%Pictures/LifetimeNaK/LLifetime.pdf}
		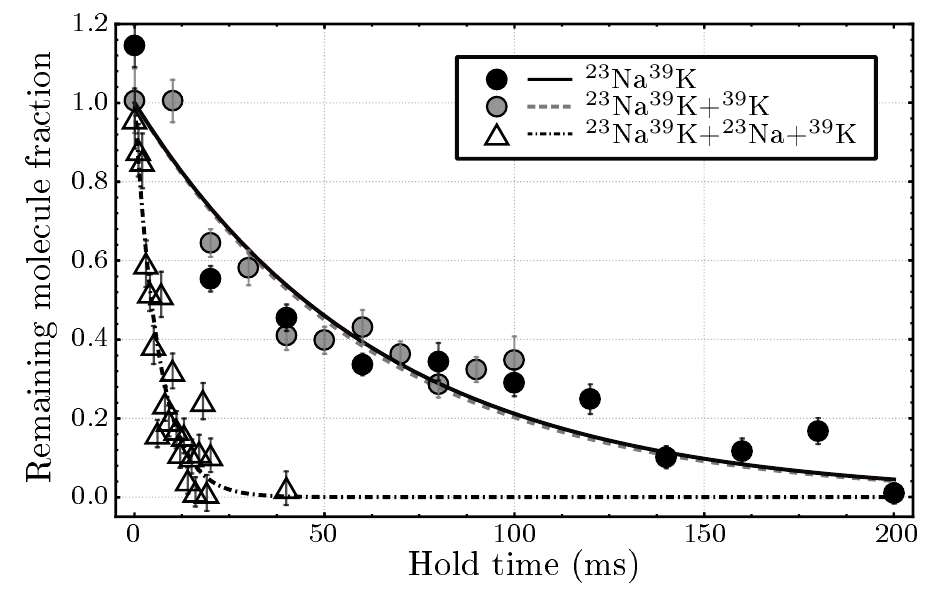}
	\caption{Loss measurements of pure ground-state molecules and with remaining atoms. The open triangles are measurements without atom removal. The fast loss originates from the chemical reaction with \Na\ atoms. The gray circles are measurements with only \Na\ removed while still \K\ atoms remain in the trap. The solid circles are measurements performed with a pure molecular ensemble. The data is normalized to the molecule number without holding time obtained from the individual fits. The curves are fits using a coupled differential equation system for modeling the losses. For the corresponding loss rate coefficients see text. All error bars are the standard deviation resulting from different experimental runs.}
	\label{LifeTimes}
\end{figure}

Heteronuclear polar ground-state molecules have attracted considerable attention in recent years. They serve as a new platform for controlled quantum chemistry \cite{Ospelkaus2010,WangCollisions}, novel many-body physics \cite{PolarMoleMoses,KRbDegenerate} and quantum simulations \cite{DeMilleQuantComp,KRbSpin}. Their permanent electric dipole moment gives rise to anisotropic and tunable long-range interactions which can be induced in the lab frame via electric fields or resonant microwave radiation \cite{KRbStereodyn,MicrowaveNa40K}. This gives exquisite control over additional quantum degrees of freedom. In recent years there has been continuous progress in the production of ultracold bialkali molecules. Fermionic $^{40}\text{K}^{87}\text{Rb}$ \cite{KRb1}, $^{23}\text{Na}^{40}\text{K}$ \cite{GsDiMo23Na40K2015} and $^{6}\text{Li}^{23}\text{Na}$ \cite{GsDiMo23Na6Li2017} as well as bosonic $^{87}\text{Rb}^{133}\text{Cs}$ \cite{GsDiMo87Rb133Cs2014Grimm} and $^{23}\text{Na}^{87}\text{Rb}$ \cite{NaRb3} molecules have been prepared.\\
Up to now, not a single molecule has been available both as a bosonic and a fermionic molecular quantum gas, which makes findings among different species and quantum statistics challenging to interpret and to compare.
For bialkali molecules only combinations with Li or K offer the possibility to prepare the bosonic and fermionic molecule, as Li and K are the only alkali metals which possess long-lived fermionic and bosonic isotopes. Among these molecules (LiK, LiNa, LiRb, LiCs, NaK, KRb, KCs) all possible combinations with a Li atom as well as the KRb molecule are known to undergo exothermic atom exchange reactions in molecule-molecule collisions \cite{Inelastic2010}. This leaves only NaK and KCs \cite{PhysRevA.95.022715} as chemically stable molecules for a comparison of scattering properties of the same molecular species but different quantum statistics.\\
Both chemically reactive and non-reactive spin-polarized fermionic molecular ensembles have been reported to be long-lived due to the centrifugal \textit{p}-wave collisional barrier limiting the two-body collisional rate to the tunneling rate \cite{GsDiMo23Na40K2015,Ospelkaus2010}. The lifetime of bosonic molecular ensembles, however, has been observed to be significantly shorter and limited by the two-body universal scattering rate \cite{NaRb3,Gregory_2019}. Two-body collisions involving molecules can lead to the formation of collisional complexes due to a large density of states. The complexes can either decay to new chemical species for chemically reactive molecules \cite{hu2019direct} or  within the lifetime of the complexes are removed from the trap by light excitation \cite{PhysRevLett.123.123402,gregory2020loss,liu2020steering} or collisions with a third scattering partner \cite{PhysRevA.87.012709,Gregory_2019}.\\
In this letter, we report on the production of ultracold bosonic \NaK\ rovibrational ground-state molecules. The preparation follows the pioneering experiments for the creation of $^{40}\text{K}^{87}\text{Rb}$ molecules \cite{KRb1} with Feshbach molecule creation and subsequent STImulated Raman Adiabatic Passage (STIRAP) transfer \cite{STIRAPBergmann2017} to a selected hyperfine state in the rovibrational ground-state manifold. We model our STIRAP transfer through an effective 5-level master equation model and work out an efficient pathway to create spin-polarized ground-state molecular ensembles. We prepare pure molecular ensembles as well as molecule-atom mixtures and extract the resulting collisional loss rate coefficients. 
We find the loss rate for the \NaK+\K\ mixture to be drastically suppressed 
which opens interesting perspectives for sympathetic cooling.\\\newline
The experiments start from ultracold weakly bound molecules. As previously described in \cite{Voges2019Fesh}, we associate \NaK\ Feshbach dimers by applying a radio frequency pulse to an ultracold mixture of bosonic \Na\ and \K\ held in a $1064\,\text{nm}$ crossed-beam optical dipole trap (cODT) with temperatures below $350\,\text{nK}$. 
We create $6\times10^3$ dimers in the least bound vibrational state $\ket{f}$ with a total angular momentum projection $M_F=-3$ and a binding energy of $h\times100\,\text{kHz}$ at a magnetic field of $199.3\,\textrm{G}$. In terms of atomic quantum numbers the state $\ket{f}$ is mainly composed of $\alpha_1\ket{m_{i,\textrm{Na}}=-3/2, m_{i,\textrm{K}}=-1/2, M_{S}=-1}+$ $\alpha_2\ket{m_{i,\textrm{Na}}=-3/2, m_{i,\textrm{K}}=-3/2, M_{S}=0}$. $M_S$ is the total electron spin projection, $m_{i,\textrm{Na/K}}$ are the nuclear spin projections and $\alpha_{1/2}$ denote the state admixtures.
For detection, we use a standard absorption technique of \K\ atoms directly from the weakly-bound molecular state.\\
For the STIRAP transfer, we make use of external-cavity diode laser systems as already described in \cite{voges2019pathway}. Both lasers are referenced simultaneously to a $10\,\text{cm}$-long high-finesse ULE cavity using a sideband Pound-Drever-Hall locking scheme \cite{Drever1983}. The cavity's finesses for the Pump and Stokes laser are 24900 and 37400, respectively, the free spectral range is $1.499\,\text{GHz}$. The linewidths of both locked lasers are estimated to be below $5\,\text{kHz}$. Furthermore, the power of the Pump laser is amplified by a tapered amplifier. Both lasers, Pump and Stokes, are overlapped and focused to the position of the molecules with $1/e^2$ Gaussian beam waists of $35$ and $40\,\mu\text{m}$, respectively. The direction of propagation is perpendicular to the direction of the magnetic field, thus $\pi$($\sigma^{+/-}$)-transitions can be addressed by choosing the polarization parallel(perpendicular) to the magnetic field.\\
\begin{figure}[t]
	\includegraphics[width=1\columnwidth]{%Pictures/STIRAPScheme/LLevelScheme.pdf}
	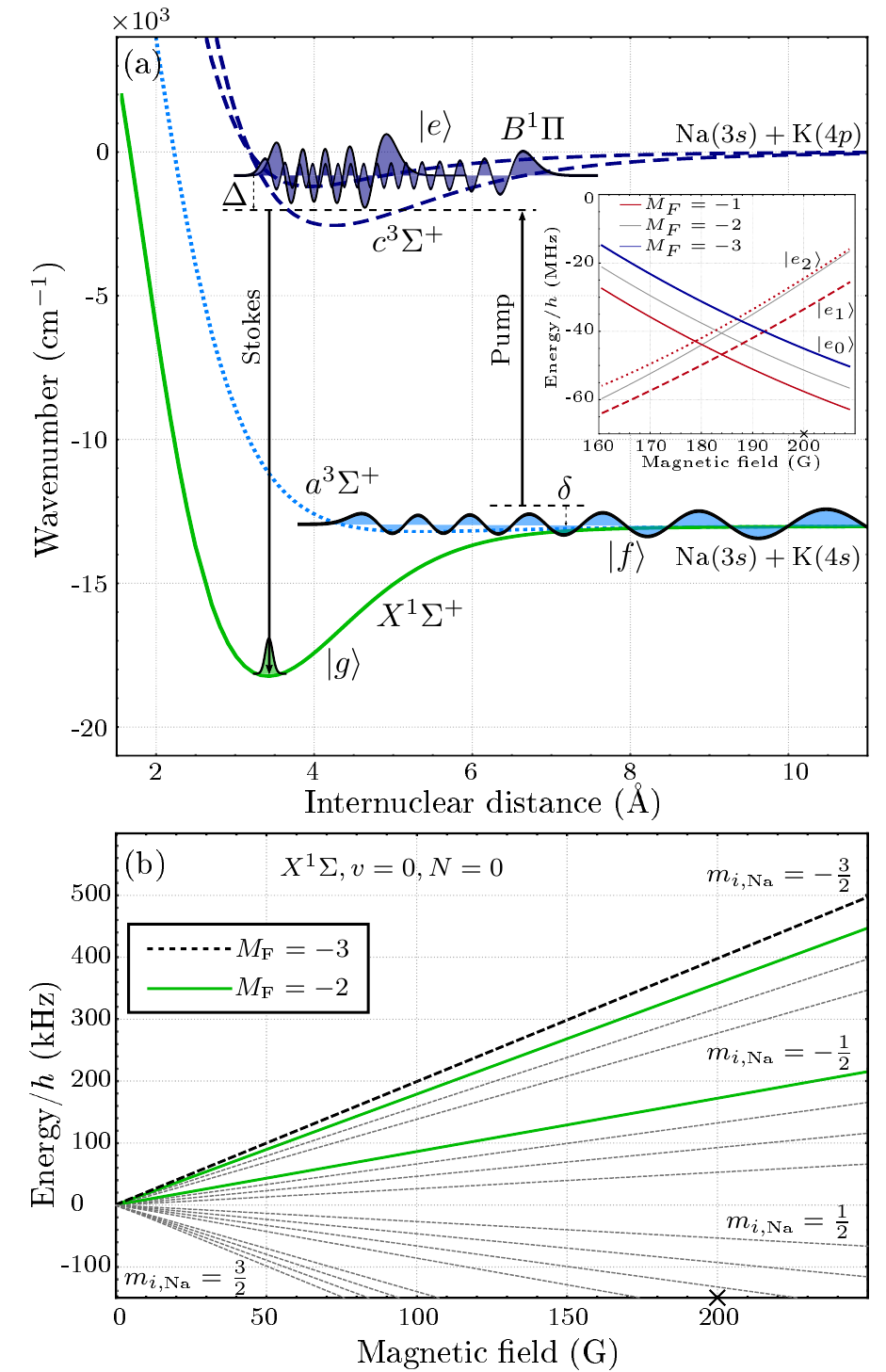}
	\caption{(a) Potential energy curves of the \NaK\ molecule. The energy is shown in $\textrm{cm}^{-1}$ as function of the internuclear distance. The solid green curve corresponds to the electronic $X^1\Sigma^+$, the dotted light blue to the $a^3\Sigma^+$ and the dashed lines to the $c^3\Sigma^+$ and $B^1\Pi$ potentials. Wavefunctions are shown as black lines with the corresponding shading. Amplitudes of the wavefunctions are not to scale. The black arrows indicate the STIRAP transitions and the one($\Delta$)- and two($\delta$)-photon detunings. The inset shows the magnetic field dependence of the Pump transition to the excited states from the model in \cite{voges2019pathway}. (b) Magnetic field dependence of the ground-state hyperfine energy structure. The green lines are the states with $M_\textrm{F}=-2$ and the black dashed line is the one with $M_\textrm{F}=-3$. As the states enter the Paschen-Back regime the four branches for different $m_{i,\textrm{Na}}$ become visible. The magnetic field, where the molecule creation process is performed, is marked with a cross on the axis.}
	\label{HFSPic}
\end{figure}\noindent
Possible transfer pathways to the ground state have been previously investigated theoretically and experimentally \cite{Schulze2013},\cite{voges2019pathway}. Figure \ref{HFSPic}(a) summarizes the relevant states involved in the transfer scheme. Starting from the weakly bound dimer state $\ket{f}$ with mainly triplet character, we make use of the triplet-singlet mixed excited state $\ket{e}$ to transfer the molecules into a selected hyperfine state in the rovibrational ground state $\ket{g}$ with pure singlet character. For the excited state $\ket{e}$ we choose the strongly spin-orbit coupled $B^1\Pi\ket{v=8}/c^3\Sigma^+\ket{v=30}$ state manifolds (see Fig.\ref{HFSPic}(a)), which have a large state admixture of $26\,\%/74\,\%$ \cite{voges2019pathway}.
The hyperfine structure of the $\ket{X^1\Sigma^+, v=0, N=0}$ ground state consists of 16 states with a total angular momentum projection $M_F=m_{i,\text{Na}}+m_{i,\text{K}}$, which group into four branches with different $m_{i,\text{Na}}$ at high magnetic fields (see Fig.\ref{HFSPic}(b)) \cite{PhysRevA.96.042506}. At $199.3\,\text{G}$, where the molecule creation is performed, the ground states are deeply in the Paschen-Back regime. In the excited states the \K\ nuclear momenta are also decoupled from the other nuclear and electronic angular momenta \cite{PhysRevA.91.032512}. Therefore, dipole transitions only change the latter ones. This limits the number of accessible ground states to three, which are highlighted in Fig.\,\ref{HFSPic}(b). Accounting only for $\pi$-transitions for the Pump transition to maximize the coupling strength, only a single state is accessible in the $c^3\Sigma^+$ hyperfine manifold, namely the $\ket{e_0}=\ket{c^3\Sigma^+, m_{i,\text{Na}}=-3/2,m_{i,\text{K}}=-1/2,M_J=-1,M_F=-3}$. The transition yields an energy of $12242.024(3)\,\text{cm}^{-1}$ (which corresponds to a wavelength of $816.8584(2)\,\textrm{nm}$) and is shown in Fig.\,\ref{HFSPic}(a). The Stokes transition, with an energy of $17453.744(3)\,\text{cm}^{-1}$ ($572.94297(10)\,\textrm{nm}$), connects the excited state to the ground state. In our case, we use a $\sigma^-$-transition to the $\ket{g}=\ket{X^1\Sigma^+, m_{i,\text{Na}}=-3/2,m_{i,\text{K}}=-1/2,M_J=0,M_i=-2}$ state. Nevertheless, our experimental setup always supports $\sigma^-$- and $\sigma^+$-transitions at the same time. Consequently, the ground state is coupled to two additional states $\ket{e_{1,2}}$ through $\sigma^+$-transitions  (see inset Fig.\,\ref{HFSPic}(a)). For the experiments and for the modeling we thus have to consider an effective 5-level system. The details of the model are described in the supplemental material.\\
  \begin{figure}[b]
	\includegraphics[width=1\columnwidth]{%Pictures/STIRAPTime/LTime.pdf}
		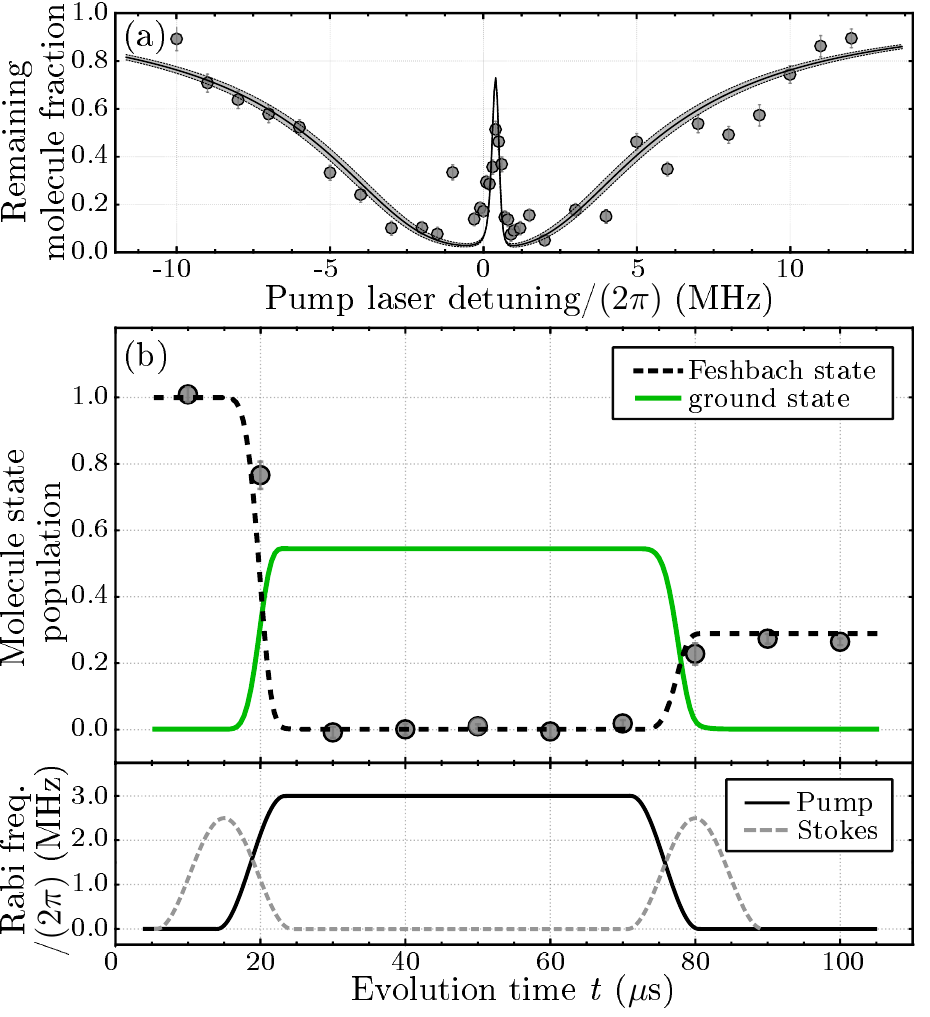}
	\caption{EIT and time evolution of STIRAP. (a) EIT measurement together with a theory curve. The points are the remaining Feshbach molecule fraction normalized to the initial Feshbach molecule number. The solid black line is the theory curve from a 3-level master equation and the dashed lines with the enclosed shaded gray area correspond to the uncertainty of the Rabi frequencies. (b) Time evolution of the Feshbach and ground-state population during a round-trip STIRAP. Data points in the upper panel are the observed Feshbach molecule number normalized to the initial molecule number. The solid green(dashed black) line is a theory curve for the ground-state(Feshbach molecule-state) population using the model described in the text and the pulses from the lower panel. The pulse duration for both lasers is $10\,\mu\textrm{s}$. The ramping up of the Pump pulse starts $1\,\mu\textrm{s}$ before the ramp down of the Stokes pulse begins. The lower panel shows the pulse sequence of the Pump and Stokes laser during the STIRAP. Rabi frequencies are obtained from one-photon loss measurements (not shown here). Error bars are the standard deviation coming from different experimental runs.}
	\label{STIRAP}
\end{figure}\newline
For STIRAP a high degree of phase coherence between the two independent laser sources is imperative. To prove the coherence and to determine the explicit frequencies for the two-photon Raman transition we perform electromagnetically induced transparency (EIT) experiments on the selected states. For the measurement shown in Fig.\,\ref{STIRAP}(a) Rabi frequencies of $\Omega_\textrm{Pump}=2\pi\times0.63(2)\,\textrm{MHz}$ and $\Omega_\textrm{Stokes}=2\pi\times4.1(2)\,\textrm{MHz}$ are used. The coherent interaction time is set to $50\,\mu\textrm{s}$. The observed asymmetry of the molecule revival arises from a one-photon detuning $\Delta=2\pi\times400(20)\,\textrm{kHz}$ to the excited state $\ket{e_0}$. EIT relies only on coherent dark state effects and never populates the ground state. A coupling between the ground state and the perturbing excited states $\ket{e_{1,2}}$ does not alter the coupling scheme as the two-photon condition is not fulfilled for these states. Consequently, a 3-level scheme is sufficient for its description.
Fig.\,\ref{STIRAP}(a) shows the experimental data and the theoretical prediction (solid black line) using experimentally determined parameters for Rabi frequencies and laser detunings. The errors on the parameters are displayed as dashed lines and gray shaded area. We find very good agreement of our data with the model and consequently good conditions for the STIRAP.\\ 
For the creation of ground-state molecules, we perform STIRAP starting from Feshbach molecules. As the Feshbach molecule lifetime is very short, on the order of 0.3\,ms \cite{Voges2019Fesh}, STIRAP is completed $25\,\mu\textrm{s}$ after Feshbach molecules creation. The STIRAP process itself takes $11\,\mu\textrm{s}$ so that no significant loss from a decay of the weakly bound dimers is expected. Figure\,\ref{STIRAP}(b) shows a typical signal for ground-state molecule creation. The figure includes the STIRAP light pulse sequence (lower panel) and the populations of the Feshbach molecules as well as the ground-state molecules during the pulse sequence calculated by a 5-level master equation. Starting with Feshbach molecules at $t=0$, the molecules are transferred to the ground state at $t= 14\,\mu\textrm{s}$ where the molecules become dark for the imaging light. To image the molecules, we reverse the STIRAP sequence and transfer ground-state molecules back to the Feshbach state. 
Due to the additional coupling of the ground state to the excited states $\ket{e_{1,2}}$, the STIRAP is highly dependent on the one-photon detuning (see Fig.\,\ref{SingPhotDetPic}). The states $\ket{e_{1,2}}$ act as loss channels, into which the ground-state molecules are pumped and consequently get lost. On resonance with one of the $\ket{e_{1,2}}$ states, nearly no ground-state molecules revive (see Fig.\,\ref{SingPhotDetPic}). Clearly, in the vicinity of the $\ket{e_{1,2}}$ states, the STIRAP benefits from fast transfers, which is restricted by the adiabaticity in the limit of small pulse-overlap areas \cite{STIRAPBergmann2017}. On the other hand, the pulse-overlap area can be increased by raising the Rabi frequencies of the pulses, which accordingly also increases the undesired coupling to the states $\ket{e_{1,2}}$. We find the best results in our system for a pulse duration of $12\,\mu\textrm{s}$ with a Pump pulse delay of $-2\,\mu\textrm{s}$ and resonant Rabi frequencies of $\Omega_\textrm{Pump}= 2\pi\times3.0(1)\,\textrm{MHz}$ and $\Omega_\textrm{Stokes}=2\pi\times2.3(1)\,\textrm{MHz}$ at a one-photon detuning $\Delta=2\pi\times8\,\textrm{MHz}$ to the center position of $\ket{e_0}$. Under these conditions single-trip STIRAP efficiency can get as high as $70\,\%$ which corresponds to a ground-state molecule number of about 4200 in a single hyperfine spin state (see inset of Fig.\,\ref{SingPhotDetPic}). Moreover, we do not observe heating effects of the molecules due to the STIRAP (see supplemental material), leading to a phase-space density of up to $0.14$. To model the influence of the states $\ket{e_{1,2}}$ on the STIRAP we apply a 5-level master equation model fit (solid curve in Fig.\,\ref{SingPhotDetPic}) and compare it to an ideal 3-level one (dashed curve in Fig.\ref{SingPhotDetPic}). The model is described in detail in the supplemental material.
In the comparison between the 5- and 3-level model the influence of the states $\ket{e_{1,2}}$ gets clear. It indicates, that the STIRAP efficiency can be easily increased by choosing a different excited state, experimental geometric condition, such as laser polarization relative to the magnetic field axis, and larger STIRAP pulse overlap areas, which is discussed in the supplemental material.\\\newline
After the transfer to the ground state the molecules are still immersed in a gas of \Na\ and \K\ atoms remaining from the creation process of the weakly bound dimers. \Na\ atoms can be removed by applying a $500\,\mu\textrm{s}$ resonant light pulse. \K\ atoms can be removed by transferring them to the $\ket{f=2,m_f=-2}$ state by a rapid adiabatic passage and a subsequent resonant light pulse for $500\,\mu\textrm{s}$. By introducing a variable hold time between the atom removals and the reversed STIRAP pulse, we perform loss measurements, which we analyze assuming a two-body decay model to extract the two-body decay rate coefficient. The model is described in the supplemental material.\\ First, we investigate the mixture of molecules and atoms. We observe fast losses from  \NaK+\Na\ collisions (see Fig.\,\ref{LifeTimes}). The extracted loss rate coefficient is $1.25(14)\times10^{-10}\,\textrm{cm}^3\textrm{s}^{-1}$, which is close to the theoretical prediction of $1.3\times10^{-10}\,\textrm{cm}^3\textrm{s}^{-1}$ taken from \cite{atoms7010036}. We assign the saturation of the losses to chemical reactions, in which $^{23}\textrm{Na}_2$ dimers are formed. Thus, \Na\ atoms are generally removed as fast as possible from the trap as the ground-state molecule number suffers from the strong losses.
\\
In a next step, we measure losses in a pure molecular ensemble (see Fig.\ref{LifeTimes}).  The two-body loss rate coefficient is measured to be $4.49(1.18)\times10^{-10}\,\textrm{cm}^3\textrm{s}^{-1}$. This loss rate coefficient is comparable to the universal limit \cite{C1CP21270B} and is possibly resulting from sticky collisions \cite{PhysRevA.87.012709} and subsequent removal of the tetramers from the trap. Comparable observations have been made in experiments with other bosonic ground-state molecules, such as $^{87}\text{Rb}^{133}\text{Cs}$ and $^{23}\text{Na}^{87}\text{Rb}$ \cite{GsDiMo87Rb133Cs2014Grimm,NaRb3}. However, the loss rate coefficient for the fermionic counterpart $^{23}\text{Na}^{40}\text{K}$ is $6\times10^{-11}\,\textrm{cm}^3\textrm{s}^{-1}$ \cite{GsDiMo23Na40K2015}. The difference can be assigned to the absence of the centrifugal barrier in bosonic \textit{s}-wave collisions.\\
Next, we investigate collisions between the molecules and \K\ atoms. Surprisingly, even a high density of \K\ atoms in the non-stretched $\ket{f=1,m_f=-1}_\textrm{K}$ state colliding with \NaK\ in the non-stretched hyperfine ground state does not increase the molecular loss (compare Fig.\,\ref{LifeTimes}), although sticky collisions with trimer formation are also expected in mixtures of \NaK+\K\ \cite{PhysRevA.85.062712}. In these collisional trimer complexes nuclear spin transitions can occur leading to subsequent loss of molecules from the prepared hyperfine state. We analyze the observed decay of the molecular cloud using the model fit described in the supplemental material. We find the loss rate coefficient for the two-body \NaK+\K\ collisions to be consistent with zero with an upper limit of $1.5\times10^{-14}\textrm{cm}^3\textrm{s}^{-1}$. The corresponding universal limit is calculated by using the prediction from \cite{PhysRevA.84.062703,PhysRevA.85.062712} and parameters from \cite{PhysRevA.82.012510} and results in $1.3\times10^{-10}\,\textrm{cm}^3\textrm{s}^{-1}$. 
Note that this corresponds to a suppression of the two-body decay in comparison to the universal limit by more than three orders of magnitude.  This is in contrast to experiments reported for fermionic molecules in collisions with bosonic atoms ($^{40}\text{K}^{87}\text{Rb}+^{87}\!\text{Rb}$\cite{Ospelkaus2010}) and fermionic atoms ($^{23}\text{Na}^{40}\text{K}+^{40}\!\text{K}$ \cite{NaKYang261}), where such suppression of losses far below the universal limit has not been observed for sticky molecule-atom collisions. The only experiment describing such a suppression has been performed in a mixture of the fermionic molecule $^{6}\text{Li}^{23}\text{Na}$ with the bosonic atom \Na\ with both particles in their lowest stretched hyperfine states \cite{son2019observation}. Here, we now report collisions in non-stretched states with loss rates far below the universal limit, which might result from a low density of resonant states \cite{PhysRevA.85.062712}. Individual resonances might thus be resolvable in this system and demand for further investigations of loss rates in other spin channels and magnetic fields.
Moreover, with the low loss rate between \NaK\ molecules and \K\ atoms in the named hyperfine state it might be possible to use \K\ atoms as a coolant for bosonic \NaK\ molecules to further increase the molecular phase-space density \cite{son2019observation}.\\\newline
In conclusion, we have reported the first creation of an ultracold high phase-space density gas of bosonic \NaK\ ground-state molecules. We have investigated the creation process and find very good agreement with our 5-level model. The spin-polarized molecular ensemble yields up to 4200 molecules and is chemically stable. We extract the two-body decay coefficient for the bosonic \NaK\ molecules. For molecule-atom collisions, we find a significant suppression of the two-body decay rate in collisions between \NaK\ molecules and \K\ atoms in non-stretched states. This unexpected result demands for further experiments including the analysis of collisions between molecules and atoms in different hyperfine states and as a function of magnetic field to identify possible scattering resonances. These experiments can be extended to a detailed comparison of collision properties between same species molecules of different quantum statistics.
\\\newline
We thank M. Siercke for enlightening comments and discussions and the group of P. O. Schmidt, PTB Brunswick, for providing scientific material for the Raman laser system. We gratefully acknowledge financial support from the European Research Council through ERC Starting Grant POLAR and from the Deutsche Forschungsgemeinschaft (DFG) through CRC 1227 (DQ-mat), project A03 and FOR 2247, project E5. P.G. thanks the DFG for financial support through RTG 1991.

\end{document}

\title{Supplemental Material: An Ultracold Gas of Bosonic \NaK\ Ground-State Molecules}

\author{Kai~K.~Voges}\email{voges@iqo.uni-hannover.de}\affiliation{Institut f\"ur Quantenoptik, Leibniz Universit\"at Hannover, 30167~Hannover, Germany}

\author{Philipp~Gersema}\affiliation{Institut f\"ur Quantenoptik, Leibniz Universit\"at Hannover, 30167~Hannover, Germany}
\author{Mara Meyer zum Alten Borgloh}\affiliation{Institut f\"ur Quantenoptik, Leibniz Universit\"at Hannover, 30167~Hannover, Germany}
\author{\\Torben A.~Schulze}\affiliation{Institut f\"ur Quantenoptik, Leibniz Universit\"at Hannover, 30167~Hannover, Germany}
\author{Torsten~Hartmann}\affiliation{Institut f\"ur Quantenoptik, Leibniz Universit\"at Hannover, 30167~Hannover, Germany}
%\author{Matthias~W.~Gempel}
%\author{Eberhard~Tiemann}
\author{Alessandro~Zenesini}\affiliation{Institut f\"ur Quantenoptik, Leibniz Universit\"at Hannover, 30167~Hannover, Germany}\affiliation{INO-CNR BEC Center and Dipartimento di Fisica, Universit\`{a} di Trento, 38123 Povo, Italy}
\author{Silke~Ospelkaus}\email{silke.ospelkaus@iqo.uni-hannover.de}\affiliation{Institut f\"ur Quantenoptik, Leibniz Universit\"at Hannover, 30167~Hannover, Germany}

\date{\today}

\begin{abstract}
In this supplement, we provide additional details on the 5-level STIRAP model, alternative STIRAP pathways for the \NaK\ molecules and the temperature measurements of the molecules. Furthermore, we detail on the loss model used for the determination of the two-body decay loss coefficients for molecule-molecule and molecule-atom collisions.
\end{abstract}
\maketitle

\subsection*{5-level STIRAP model}
STImulated Raman Adiabatic Passage (STIRAP) for the transfer of weakly bound Feshbach molecules to the ground state, and vice versa, is typically performed in a pure 3-level $\Lambda$-system \cite{STIRAPBergmann2017}. In our case, the Feshbach molecule state is named $\ket{f}$, the ground state $\ket{g}$ and the excited states are named $\ket{e_i}$. The laser beams for the Pump and the Stokes transitions are copropagating and perpendicular to the magnetic field. For both beams, linear polarizations parallel ($\parallel$) to the magnetic field access $\pi$-transitions in the molecules and linear polarizations perpendicular ($\perp$) to the magnetic field access always both $\sigma^+\,\textrm{- and}\ \sigma^-$-transitions.\\
The molecular starting state $\ket{f}$ can be described as a composed state of $\alpha_1\ket{m_{i,\textrm{Na}}=-3/2,m_{i,\textrm{K}}=-1/2,M_{S}=-1}+$ $\alpha_2\ket{m_{i,\textrm{Na}}=-3/2,m_{i,\textrm{K}}=-3/2, M_{S}=0}$, where $M_S$ is the total electron spin projection and $\alpha_{1/2}$ represent state admixtures. With the goal of maximizing the Rabi frequency $\Omega_\textrm{P(ump)}$, we choose excited states from the triplet hyperfine manifold of the coupled triplet-singlet states $\ket{c^3\Sigma^+, v=30}$ and $\ket{B^1\Pi,v=8}$ \cite{Schulze2013,voges2019pathway}. Moreover, we choose the polarization of the Pump beam to be $\parallel$. The only possible accessible excited state is the  $\ket{e_0}=\ket{m_{i,\text{Na}}=-3/2,m_{i,\text{K}}=-1/2,M_J=-1,M_F=-3}$. Using $\perp$ polarization for the Stokes laser, we reach the $\ket{g}=\ket{m_{i,\text{Na}}=-3/2,m_{i,\text{K}}=-1/2,M_J=0,M_i=-2}$ ground state with a $\sigma^-$-transition. Other states in the ground state cannot be reached, because the ground state manifold has pure singlet character and is deeply in the Paschen-Back regime. Thus, nuclear and electronic spins are decoupled so that only the electronic spin projection can be changed by an optical transition.\\
At the same time $\sigma^+$-transitions couple the state $\ket{g}$ to the excited state $\ket{e_{1,2}}$ which have both state contributions in the atomic base from $\ket{m_{i,\text{Na}}=-3/2,m_{i,\text{K}}=-1/2,M_J=1,M_F=-1}$. Note that the Pump beam does not couple the state $\ket{f}$ to the states $\ket{e_{1,2}}$ due to $\Delta M_F=2$.\\
In summary, the experimental situation requires to extend the typical 3-level $\Lambda$-system (for the state $\ket{f}$, $\ket{e_0}$ and $\ket{g}$) to a 5-level system (for the states $\ket{f}$, $\ket{e_0}$, $\ket{g}$ and $\ket{e_{1,2}}$). The model Hamilton operator $H(t)$ for the light-molecule interaction and the molecular energies in the rotating-wave-approximation is
\begin{equation*}
\hbar\begin{bmatrix}
	0&\Omega_\textrm{P}(t)/2&0&0&0\\
	\Omega_\textrm{P}(t)/2&\Delta_\textrm{P}&\Omega_\textrm{S}(t)/2&0&0\\
	0&\Omega_\textrm{S}(t)/2&\Delta_\textrm{P}-\Delta_\textrm{S}&\Omega_\textrm{S,1}(t)/2&\Omega_\textrm{S,2}(t)/2\\
	0&0&\Omega_\textrm{S,1}(t)/2&\Delta_\textrm{P}-\Delta_\textrm{S,1}&0\\
	0&0&\Omega_\textrm{S,2}(t)/2&0&\Delta_\textrm{P}-\Delta_\textrm{S,2}\\
\end{bmatrix}.
\end{equation*}
 The time dependent state vector is represented by $(c_f(t),c_{e_0}(t),c_g(t),c_{e_1}(t),c_{e_2}(t))^\mathrm{T}$, where $c_i$ is the probability amplitude of the corresponding state $\ket{i}$. $\Omega_\textrm{P}(t)$ is the Rabi frequency for the Pump transition and  $\Omega_\textrm{S(tokes)}(t)$, $\Omega_\textrm{S,1}(t)$ and $\Omega_\textrm{S,2}(t)$ are the Rabi frequencies for the Stokes transition to the excited states $\ket{e_0}$,$\ket{e_1}$ and $\ket{e_2}$, respectively. Note, that all Rabi frequencies are time dependent and real. $\Delta_\textrm{P}$ and $\Delta_\textrm{S}$ are the detunings of the Pump and Stokes laser frequency to the respective molecular transition. The relative positions of the excited states $\ket{e_{1,2}}$ to $\ket{e_0}$ are $\Delta_\textrm{S,1}=2\pi\times(-10)\,\textrm{MHz}$ and $\Delta_\textrm{S,2}=2\pi\times(-21)\,\textrm{MHz}$, respectively, at 199.3\,G and are taken from our excited state model presented in \cite{voges2019pathway}. To additionally model losses of the molecules from the excited states, a sixth state $\ket{l}$ is introduced, which is not directly coupled to any other state. This is important for the numerical calculation, as it keeps the population normalized during the evaluation. The dynamics of the system can be modeled by solving the master equation in Lindblad representation with the density matrix $\rho(t)$
\begin{equation}
\dot\rho(t)=-\frac{i}{\hbar}[ H(t),\rho(t)]+\sum_k\gamma_k D[A_k]\rho(t)\,.
\end{equation}
The second term denotes the losses from the system, where $\gamma_k$ are the decay rates of the excited states which we set for all three states to $\gamma_0=\gamma_1=\gamma_2=2\pi\times 11\,\textrm{MHz}$ and $D[A_k]$ are the corresponding Lindblad superoperators with the jump operator $A_k$ from the excited state $\ket{e_k}$ to the loss state $\ket{l}$ \cite{CarmichaelBook}.\\
For the fit of the experimental data in Fig.\,3 we use this model with the Rabi frequencies $\Omega_\textrm{S,1}$ and $\Omega_\textrm{S,2}$ as free parameters as well as the STIRAP Rabi frequencies $\Omega_\textrm{P}$ and $\Omega_\textrm{S}$  constrained to their experimentally determined uncertainties. We assign the optimum of the fit within these constrains, confirming the consistency of our data. Furthermore, this model was used to also calculate the STIRAP time dynamics of Fig.\,2(b).\\
The 5-level model can be reduced to a 3-level one by setting the coupling to the excited state $\ket{e_{1,2}}$ to zero. We use this to calculate the theoretical electromagnetically induced transparency curve in Fig.\,2(a) and the optimal curve for the one-photon detuning (gray dashed line) in Fig.\,3.
\subsection*{Alternative STIRAP pathways}
\begin{figure}[t]
  	
	\includegraphics[width=1\columnwidth]{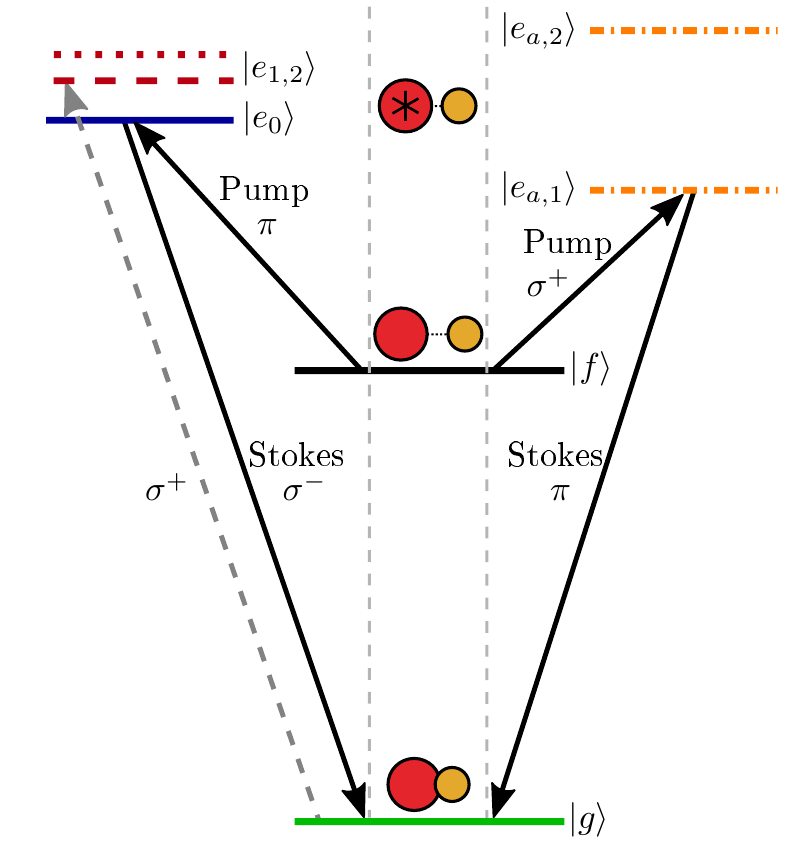}
	\caption{STIRAP pathways. This figure shows the current (left) and the alternative (right) STIRAP pathway. Pathways start from the Feshbach molecule state $\ket{f}$ (black solid line) and end in the ground state $\ket{g}$ (green solid line). The excited states $\ket{e_i}$ are the ones described in the text. For the current STIRAP the Pump beam drives $\pi$-transitions and the Stokes beam $\sigma^-$-transitions, displayed as solid arrows. On contrary, the Stokes beam couples also with $\sigma^+$-transition to the excited states $\ket{e_{1,2}}$(dashed arrow). The alternative STIRAP pathways use  $\sigma^+$-transitions for the Pump and $\pi$-transitions for the Stokes beam. The states $\ket{e_{a,1;2}}$ are shown as orange dashed-dotted lines.}
	\label{NewSTIRAP}
\end{figure}

Alternative STIRAP pathways using states from the $c^3\Sigma^+$ potential may be possible with either another STIRAP beam alignment, for example parallel to the magnetic field direction, and/or with other polarizations. In case of a perpendicular alignment, as it is described above, alternative STIRAP pathways to the ground state $\ket{g}$ are possible when switching the laser polarizations, using now $\perp$ polarization for the Pump and $\parallel$ polarization for the Stokes beam; see Fig.\ref{NewSTIRAP}. 
  \newline
We identify two additional states $\ket{e_{\textrm{a,1}}}$ and $\ket{e_{\textrm{a,2}}}$ suiting these pathways, both yielding state contributions from the $\ket{m_{i,\text{Na}}=-3/2,m_{i,\text{K}}=-1/2,M_J=0,M_F=-2}$ in the atomic base. Their transitions are $+189$ and $-146$\,MHz detuned from the original one $\ket{e_0}$ and do not possess neighboring states close by which may be populated through $\sigma^-$-transitions to the state $\ket{f}$. The additional STIRAP pathways are identified based on the model of the excited states \cite{voges2019pathway}. Simulations, utilizing the model described above suggest round-trip efficiencies of more than 80\,\%. These states will be object of future investigation.

\subsection*{Loss coefficients}
Two-body loss coefficients for molecule-molecule and molecule-atom collisions are extracted from the decay of the \NaK\ ground-state molecule ensemble.\\
In a pure molecular ensemble, losses can be assigned to two-body losses with tetramer formation and subsequent removal or loss of the tetramers, see \cite{Gregory_2019}. We obtain an analytic solution for the two-body loss of the ground-state molecule number $N_\textrm{NaK}(t)$ \cite{NoteAle}
\begin{equation}
	N_\textrm{NaK}(t)=\frac{N_{\textrm{NaK},0}}{(1+\frac{11}{8}\,\epsilon\,k_{\textrm{NaK},2} t)^{8/11}}\,,\label{eq:twobody}
\end{equation}
where $N_{\textrm{NaK},0}$ is the initial ground-state molecule number, $k_{\textrm{NaK},2}$ the molecular two-body loss coefficient and $\epsilon=(m_\textrm{NaK}\bar\omega/(2\pi k_\textrm{B}))^{3/2}$ with $\bar\omega$ the average trap frequency and $k_\textrm{B}$ the Boltzmann constant. $N_{\textrm{NaK},0}$ and $k_{\textrm{NaK},2}$ were used as free parameters for the fit.\\
For the model of the loss from molecule-atom collisions we use the coupled differential equation system:
\begin{align}
	\dot N_\textrm{NaK}(t)&=-\epsilon\,k_{\textrm{NaK},2}\frac{N_\textrm{NaK}(t)^2}{T_\textrm{NaK}(t)^{3/2}}- \eta\,k_\textrm{a}N_{\textrm{a},0}N_\textrm{NaK}(t) \nonumber \\
	\dot T_\textrm{NaK}(t)&=\epsilon\,k_{\textrm{NaK},2}\frac{N_\textrm{NaK}(t)}{4\sqrt{T_\textrm{NaK}(t)}}\,.\label{eq:moleloss}
\end{align}
$\eta$ is the density overlap between molecules and atoms, $k_\textrm{a}$ the loss coefficient for the molecule-atom collision, $N_{\textrm{a},0}$ the initial atom number and $T_\textrm{NaK}$ the temperature of the ground-state molecules.\\
Note, that in the model anti-evaporation effects may be considered for molecules only, or for both, molecules and atoms. The difference of these two cases is smaller than our experimental uncertainties. The presented data only consider effects on molecules.
\subsection*{Temperature measurement for molecular clouds}
Temperature measurements for atomic clouds are typically done through time-of-flight (TOF) measurements after releasing them from the trap and fitting a temperature dependent expansion curve to the clouds width. For ground-state molecules, this technique is limited by the free expansion time, as the molecules might leave the region of the STIRAP laser beams which is needed to transfer the ground-state molecules back to the Feshbach state for imaging. In our experiment, the STIRAP beam foci have $1/e^2$ radii of 35 and 40\,$\mu$m, respectively, allowing for almost no free expansion time of the molecules before leaving the STIRAP beam area.\\
To still measure the temperature of the ground-state molecule ensemble we reverse the entire molecule creation process, by means of STIRAP and Feshbach molecule dissociation, before performing the TOF and imaging on the dissociated atoms. Note, that for our Feshbach molecules, imaging normally takes place from the Feshbach molecule state itself, as the linewidth of the imaging transition in \K\ is larger than the binding energy of the weakly bound dimers \cite{Voges2019Fesh}. A temperature measurement in TOF with Feshbach molecules is also not possible, because the Feshbach molecule lifetime is very short, about 1\,ms in a pure molecular ensemble, and an appropriate signal would be lost very fast. A complete dissociation within $600\,\mu\textrm{s}$ using a resonant radio frequency of $210.0\,\textrm{MHz}$ to the $\ket{f=1,m_f=-1}_\textrm{Na}+\ket{f=1,m_f=-1}_\textrm{K}$ is performed immediately after the backwards STIRAP. Temperature TOF measurements are then performed on the long living atomic ensemble. All atoms involved in the temperature measurement come originally from deeply bound molecules. \\
The extracted temperature from the atoms show the same temperature as the initial atoms measured before the molecule creation happened. Consequently, all transfers in between (Feshbach molecule creation, STIRAPs, atom state preparations and removals, Feshbach molecule dissociation) do not heat the molecule ensemble.